\RequirePackage{fixltx2e}
\documentclass[oneside,english]{amsart}
\usepackage[T1]{fontenc}
\usepackage[latin9]{inputenc}
\usepackage{babel}
\usepackage{amsthm}
\usepackage{graphicx}
\usepackage{setspace}
\usepackage[unicode=true,pdfusetitle,
 bookmarks=true,bookmarksnumbered=false,bookmarksopen=false,
 breaklinks=false,pdfborder={0 0 1},backref=false,colorlinks=false]
 {hyperref}

\makeatletter
\numberwithin{equation}{section}
\numberwithin{figure}{section}
\numberwithin{table}{section}
\theoremstyle{plain}
\newtheorem{thm}{\protect\theoremname}[section]
\theoremstyle{definition}
\newtheorem{defn}[thm]{\protect\definitionname}
\theoremstyle{plain}
\newtheorem{cor}[thm]{\protect\corollaryname}

\makeatother

\providecommand{\corollaryname}{Corollary}
\providecommand{\definitionname}{Definition}
\providecommand{\theoremname}{Theorem}

\begin{document}

\title{Circuit complexity and Problem structure in Hamming space}

\author{Koji KOBAYASHI}

\date{2018-05-29}
\begin{abstract}
This paper describes about relation between circuit complexity and
accept inputs structure in Hamming space by using almost all monotone
circuit that emulate deterministic Turing machine (DTM). 

Circuit family that emulate DTM are almost all monotone circuit family
except some NOT-gate which connect input variables (like negation
normal form (NNF)). Therefore, we can analyze DTM limitation by using
this NNF Circuit family.

NNF circuit have symmetry of OR-gate input line, so NNF circuit cannot
identify from OR-gate output line which of OR-gate input line is 1.
So NNF circuit family cannot compute sandwich structure effectively
(Sandwich structure is two accept inputs that sandwich reject inputs
in Hamming space). NNF circuit have to use unique AND-gate to identify
each different vector of sandwich structure. That is, we can measure
problem complexity by counting different vectors.

Some decision problem have characteristic in sandwich structure. Different
vectors of Negate HornSAT problem are at most constant length because
we can delete constant part of each negative literal in Horn clauses
by using definite clauses. Therefore, number of these different vector
is at most polynomial size. The other hand, we can design high complexity
problem with almost perfct nonlinear (APN) function.
\end{abstract}

\maketitle

\section{Introduction}

In this paper, we consider the relation between circuit complexity
and accept inputs structure in Hamming space by using almost all monotone
circuit that emulate deterministic Turing machine (DTM). For example,
we analyze Negation HornSAT problem complexity and design new problem
that include high complexity of primitive polynomial non linearity.

In computational complexity, we use circuit family to analyze problem
complexity, and we find out some result such as $PARITY\notin AC_{0}$
\cite{Ajtai,Furst}, $CLIQUE\notin mP$ monotone circuit family with
polynomial size \cite{Razborov}. The purpose of this paper is to
provide new approach to analyze problem complexity by corresponding
problem input structure in Hamming space and gate in circuit family
which emulate DTM.

\section{NNF circuit family}

First, we define NNF circuit family that is almost all monotone circuit.
Explained in book \cite{Sipser} Circuit Complexity section 9.30,
Circuit family can emulate DTM  only using NOT-gate in changing input
values $\left\{ 0,1\right\} $ to $\left\{ 01,10\right\} $. This
``almost all monotone circuit family'' have simple structure like
monotone circuit family.
\begin{defn}
\label{def: NNF Circuit family}\textcompwordmark{}

We will use the terms;

``NNF Circuit Family'' as circuit family that have no NOT-gate except
connecting INPUT-gates directly (like negation normal form).

``Input variable pair'' as output pair of INPUT-gate and NOT-gate
$\left\{ 01,10\right\} $ that correspond to an input variable $\left\{ 0,1\right\} $.
\end{defn}

\begin{figure}
\begin{centering}
\includegraphics{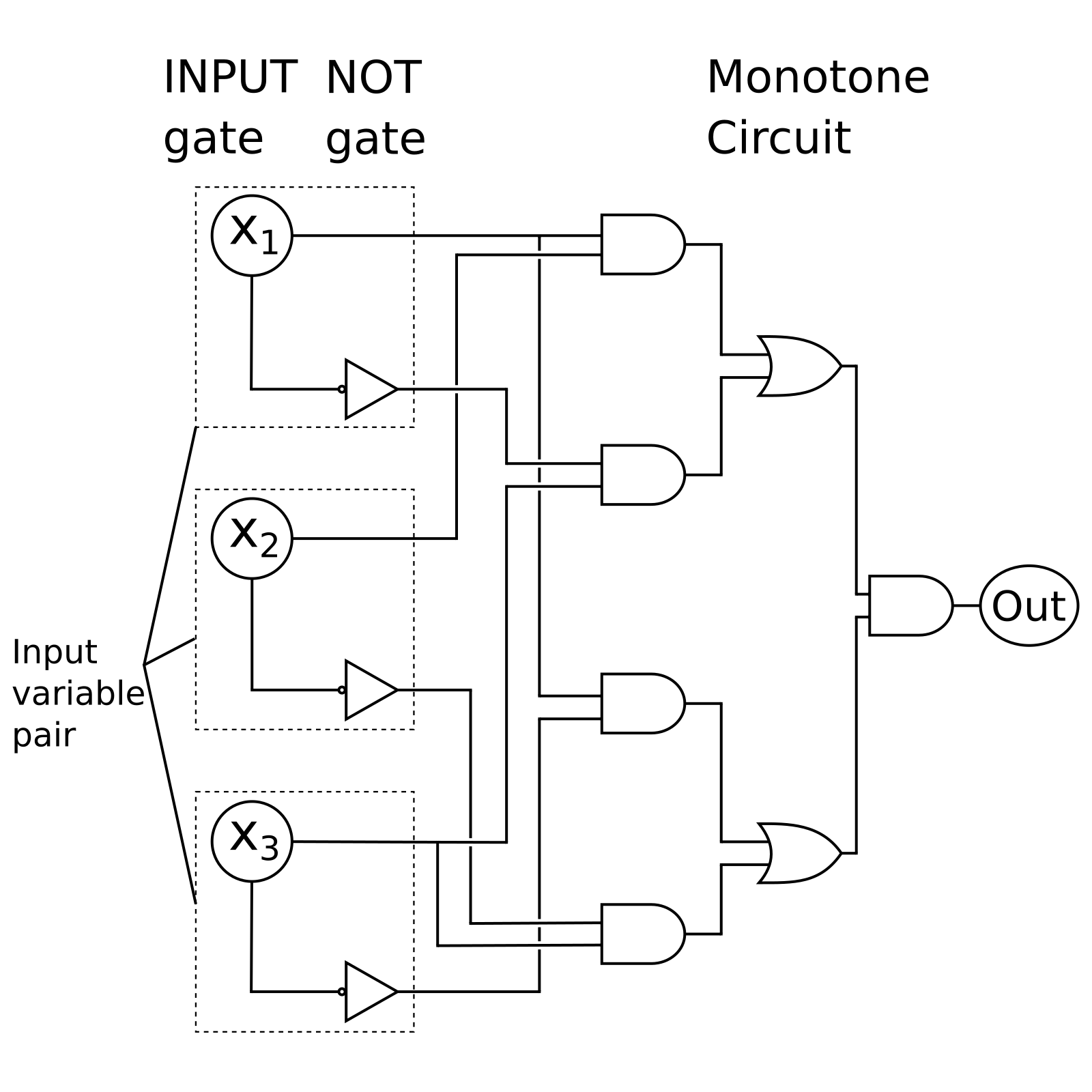}
\par\end{centering}
\caption{\label{fig: NNF Circuit}NNF circuit}
\end{figure}

Figure \ref{fig: NNF Circuit} is example of a NNF circuit.

\begin{thm}
\label{thm: Emulate DTM by using NNF Circuit family}\textcompwordmark{}

Let $t:N\longrightarrow N$ be a function where $t\left(n\right)\geq n$.

If $A\in TIME\left(t\left(n\right)\right)$ then NNF circuit family
can emulate DTM that compute $A$ with $O\left(t^{2}\left(n\right)\right)$
gate.
\end{thm}

\begin{proof}
This Proof is based on \cite{Sipser} theorem 9.30 proof. See \cite{Sipser}
for detail.

NNF circuit family can emulate DTM by computing every step's cell
values (and head state if head on the cell). Figure \ref{fig: NNF Circuit Block}
shows part of a NNF circuit block diagram.

\begin{figure}
\begin{centering}
\includegraphics{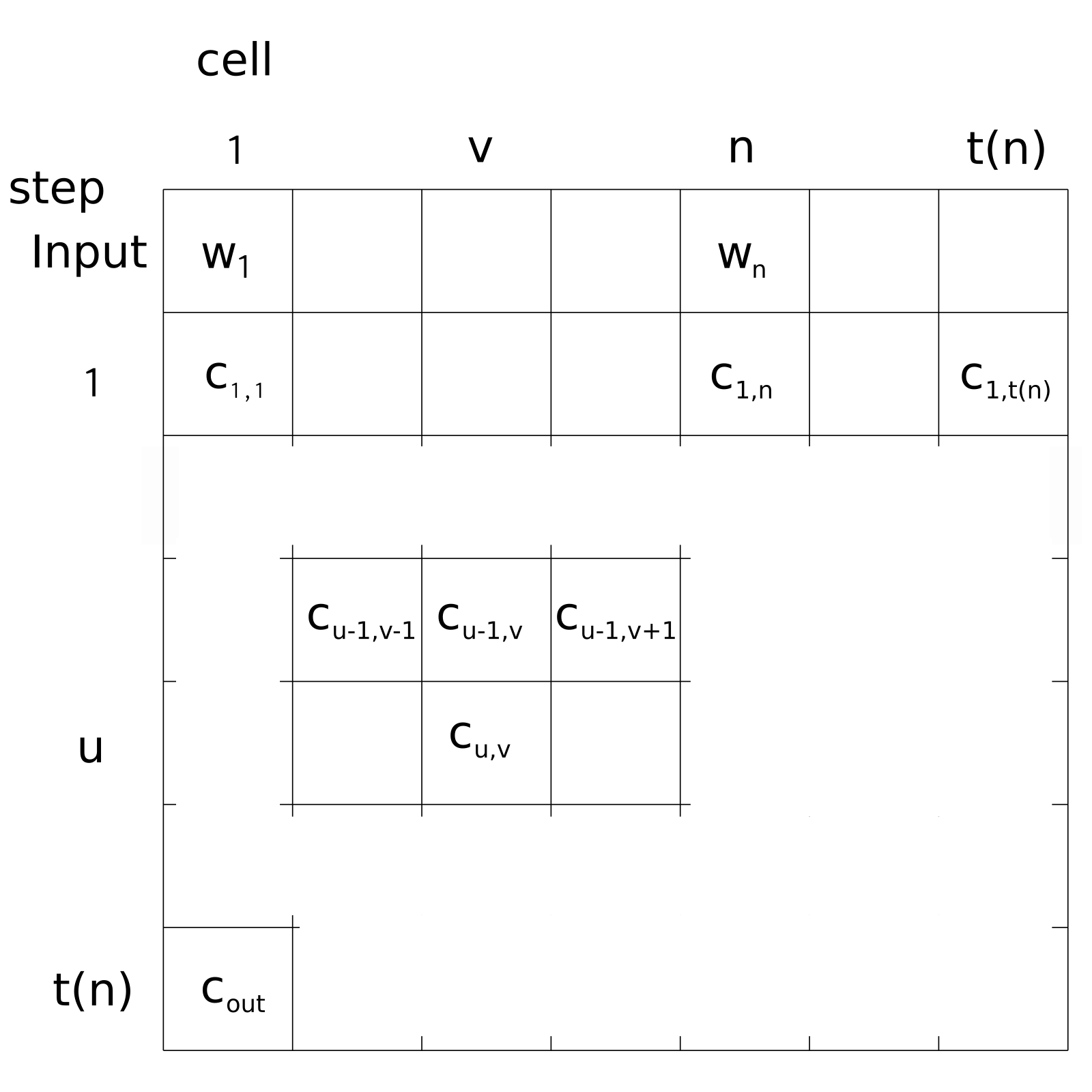}
\par\end{centering}
\caption{\label{fig: NNF Circuit Block}NNF circuit block diagram}
\end{figure}

Input of this circuit is modified $w_{1}\cdots w_{n}$ to $c_{1,1}\cdots c_{1,n}$,
and finally output result at $c_{out}=c_{t\left(n\right),1}$ cell.
This circuit emulate DTM behavior, so $c_{u,v}$ compute cell's state
of step $u$ from previous step cell $c_{u-1,v}$ and each side cells
$c_{u-1,v-1},c_{u-1,v+1}$ (because head affect at most side cells
in each step).

Figure \ref{fig: Cuv Circuit} shows example of $c_{u,v}$ sub circuit
that transition function is ``if state is $q_{k}$ and tape value
is $0$, then move $+1$ and change state to $q_{m}$''. This circuit
shows one of transition configuration which $\left(c_{u-1,v-1},c_{u-1,v},c_{u-1,v+1}\right)=\left(q_{k}0,q_{-}0,q_{-}0\right)$.
$q_{-}$ means ``no head on the cell''. 

\begin{figure}
\begin{centering}
\includegraphics{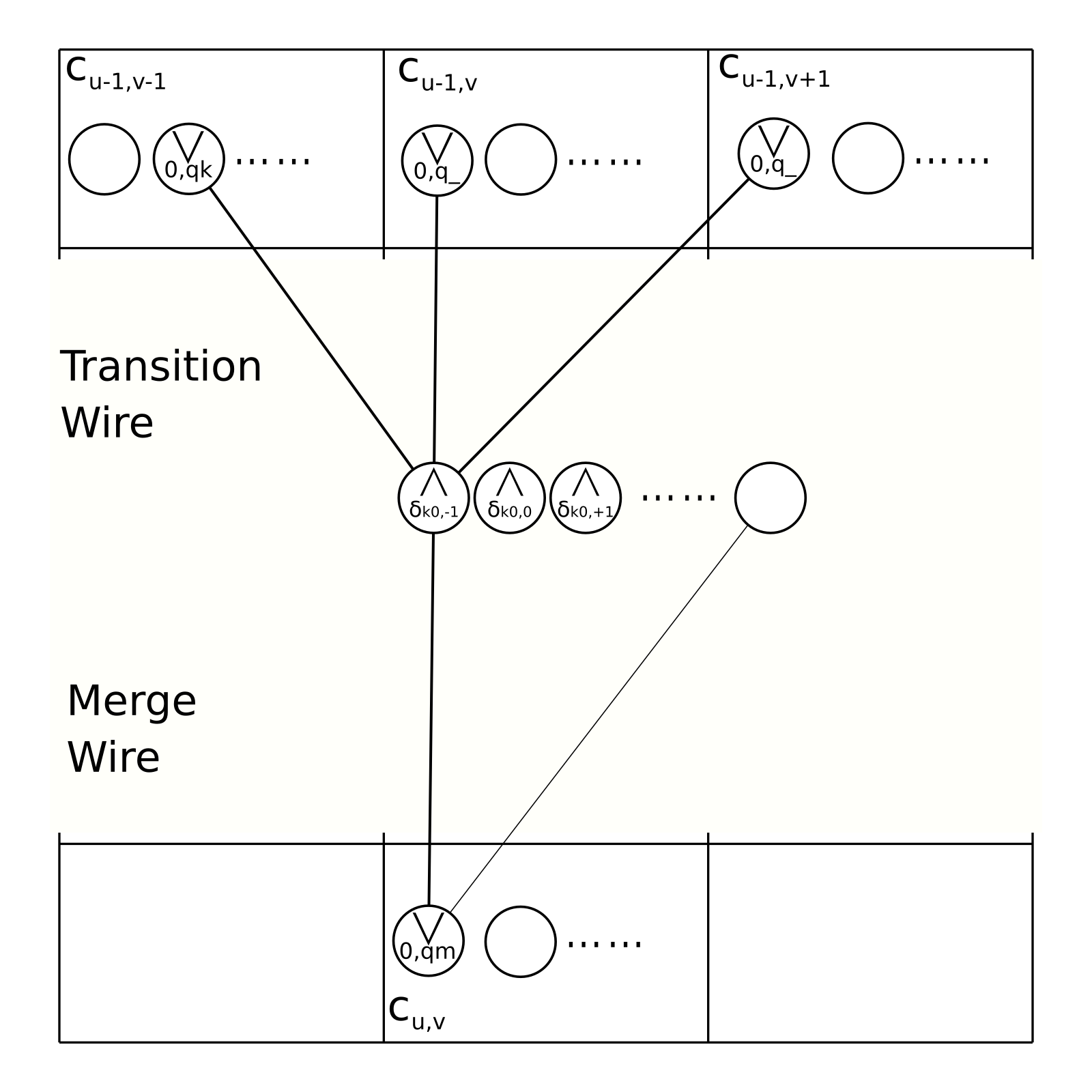}
\par\end{centering}
\caption{\label{fig: Cuv Circuit}$c_{u,v}$ circuit}
\end{figure}

Each OR-gate $\vee_{w,q}$ in $c_{u,v}$ correspond to every step's
cell condition (cell value $w$, and head status $q$ if head exist
on the $c_{u,v}$ cell), and output 1 if and only if $c_{u,v}$ cell
satisfy corresponding condition. Previous step's $\vee$ output in
$c_{u-1,v-1}$, $c_{u-1,v}$, $c_{u-1,v+1}$ are connected to next
step's AND-gate $\wedge_{\delta}$ in $c_{u,v}$ with transition wire.
Each $\wedge_{\delta}$ correspond to transition function $\delta$,
and each $\wedge_{\delta}$ output correspond to each transition function's
result of $c_{u,v}$. To simplify, NNF circuit include separate three
gates $\wedge_{\delta,-1}$, $\wedge_{\delta,0}$, $\wedge_{\delta,+1}$
according to head exists position $c_{u-1,v-1}$, $c_{u-1,v}$, $c_{u-1,v+1}$,
and special transition function $\delta_{-}$ which correspond to
no head transition (keep current tape value). So $\wedge_{\delta}$
in $c_{u,v}$ output 1 if and only if previous step's $\vee$ output
in $c_{u-1,v-1}$, $c_{u-1,v}$, $c_{u-1,v+1}$ satisfy transition
function $\delta$ condition. Each transition functions affect (or
do not affect) next step's condition, so $\wedge_{\delta}$ output
is connected to $\vee_{w,qm}$ in $c_{u,v}$ and decide $c_{u,v}$
condition. Because DTM have constant number of transition functions,
NNF can compute each step's cell by using constant number of AND-gates
and OR-gates (without NOT-gate).

First step's cells are handled in a special way. Input is $\left\{ 0,1\right\} ^{*}$
and above monotone circuit cannot manage $0$ value. So NNF circuit
compute $\left\{ 0,1\right\} ^{*}\longrightarrow\left\{ 01,10\right\} ^{*}$
by using NOT-gate.

\begin{figure}
\begin{centering}
\includegraphics{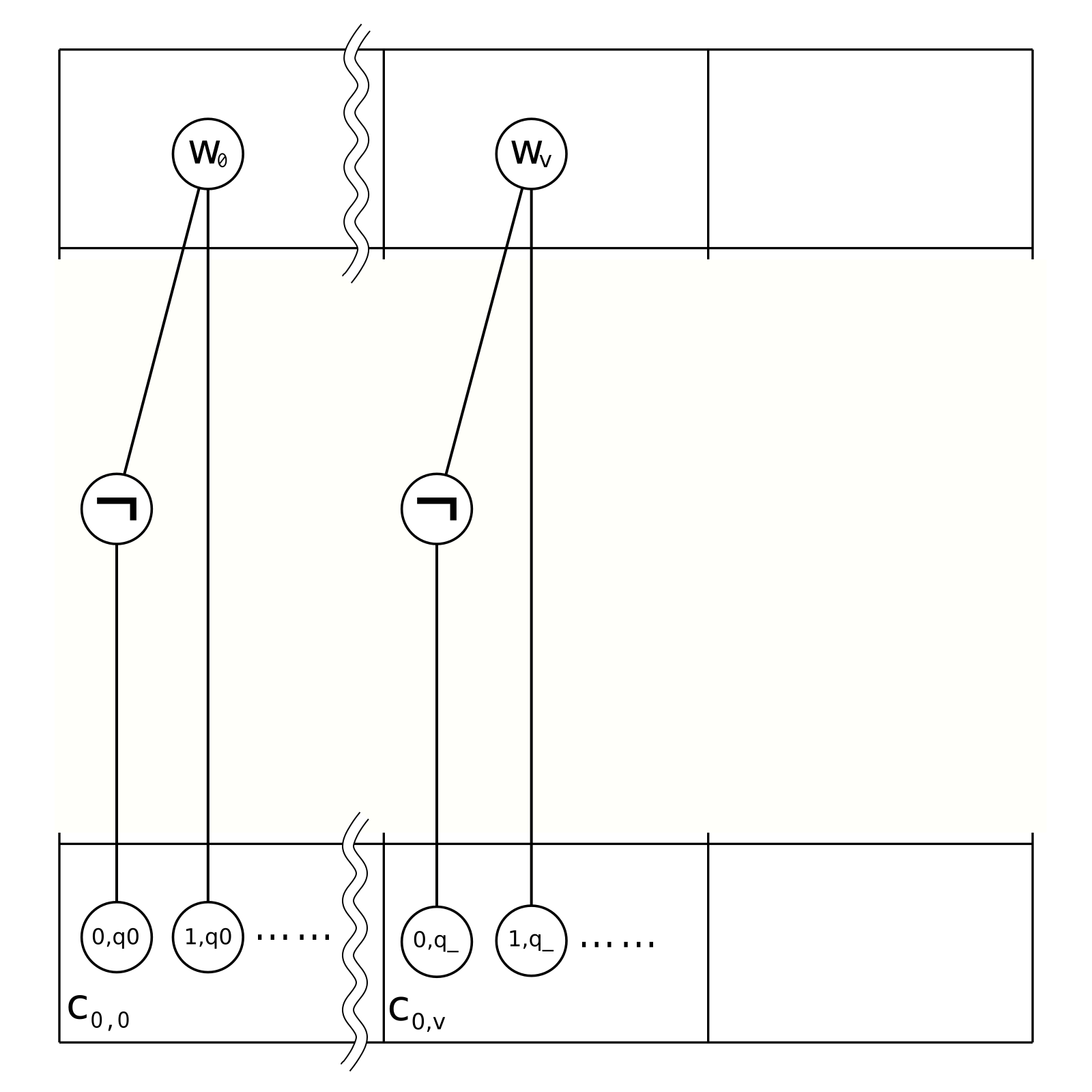}
\par\end{centering}
\caption{\label{fig: First step} First step}
\end{figure}
\end{proof}
\begin{cor}
\label{cor: NNF circuit family and P}\textcompwordmark{}

NNF circuit family can compute P problem with polynomial number of
gates of input length.
\end{cor}

Confirm NNF circuit family behavior. We define some term that decide
relation of inputs.
\begin{defn}
\label{def: Neighbor input space}\textcompwordmark{}

We will use the term;

``Neighbor input (pair)'' as accept inputs pair that no accept inputs
exists between these accept input in Hamming space.

``Boundary input (set) of neighbor input'' as reject inputs that
exist between neighbor inputs in Hamming space.

``Different variables'' as all difference part of values in neighbor
input pair.

``Different vector'' as vector and inverse vector pair which start
and end point is neighbor input pair in Hamming space. To simplify,
we use $\overline{1}=-1$.

``Neighbor distance'' as different vector length.

``Sandwich structure'' as connected graph which nodes are accept
inputs in Hamming space.

Figure \ref{fig: Sandwich structure} shows example of sandwich structure
which neighbor input pair is $0000111110011000$ and $0000000000000000$.
In this case, $\_\_\_\_11111\_\_11\_\_\_$ and $\_\_\_\_00000\_\_00\_\_\_$
are different variables, and $\left(0000111110011000\right)$ and
$\left(0000\overline{1}\overline{1}\overline{1}\overline{1}\overline{1}00\overline{1}\overline{1}000\right)$
are different vector, neighbor distance is 7.

\begin{figure}
\begin{centering}
\includegraphics{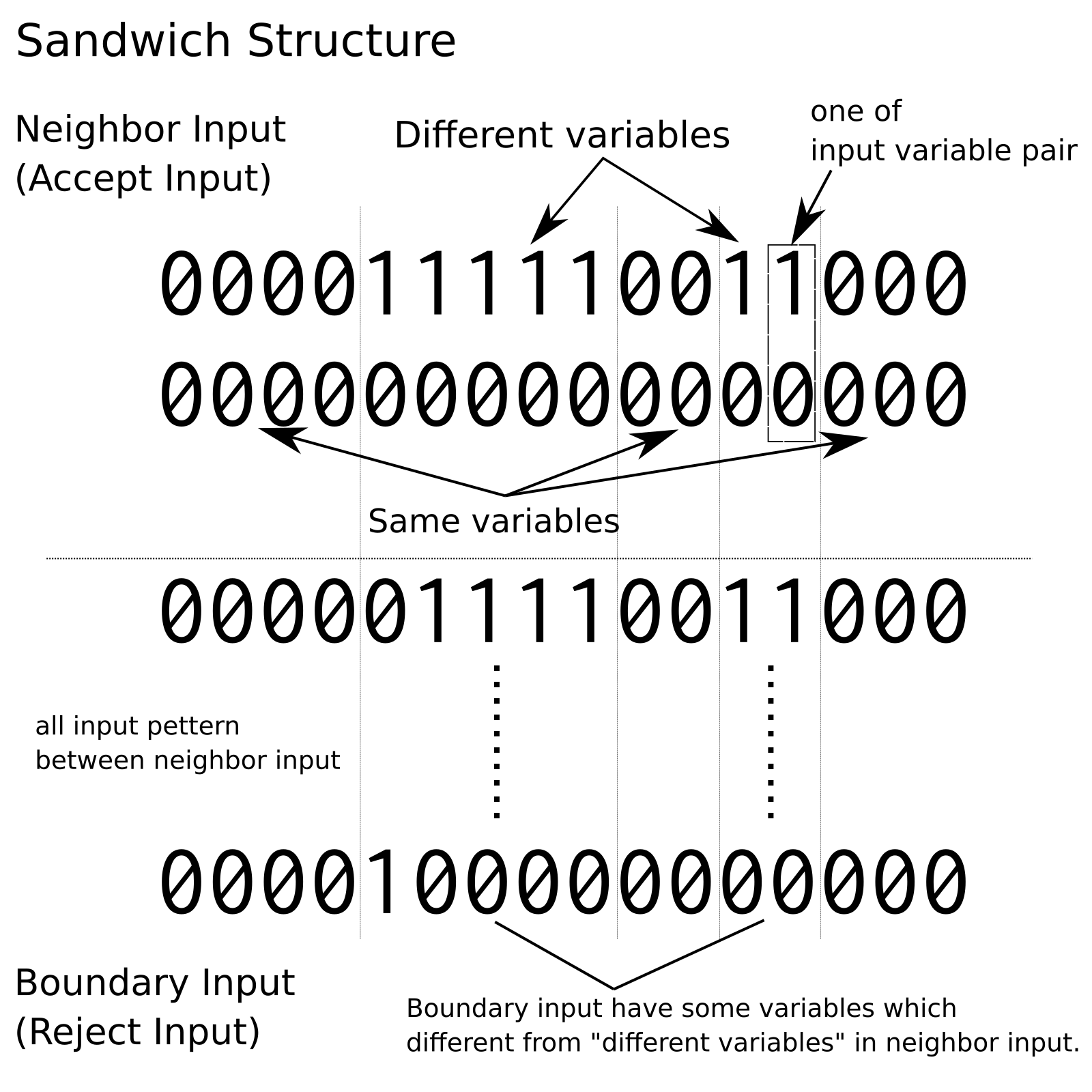}
\par\end{centering}
\caption{\label{fig: Sandwich structure} Sandwich structure}
\end{figure}

\begin{figure}
\begin{centering}
\includegraphics{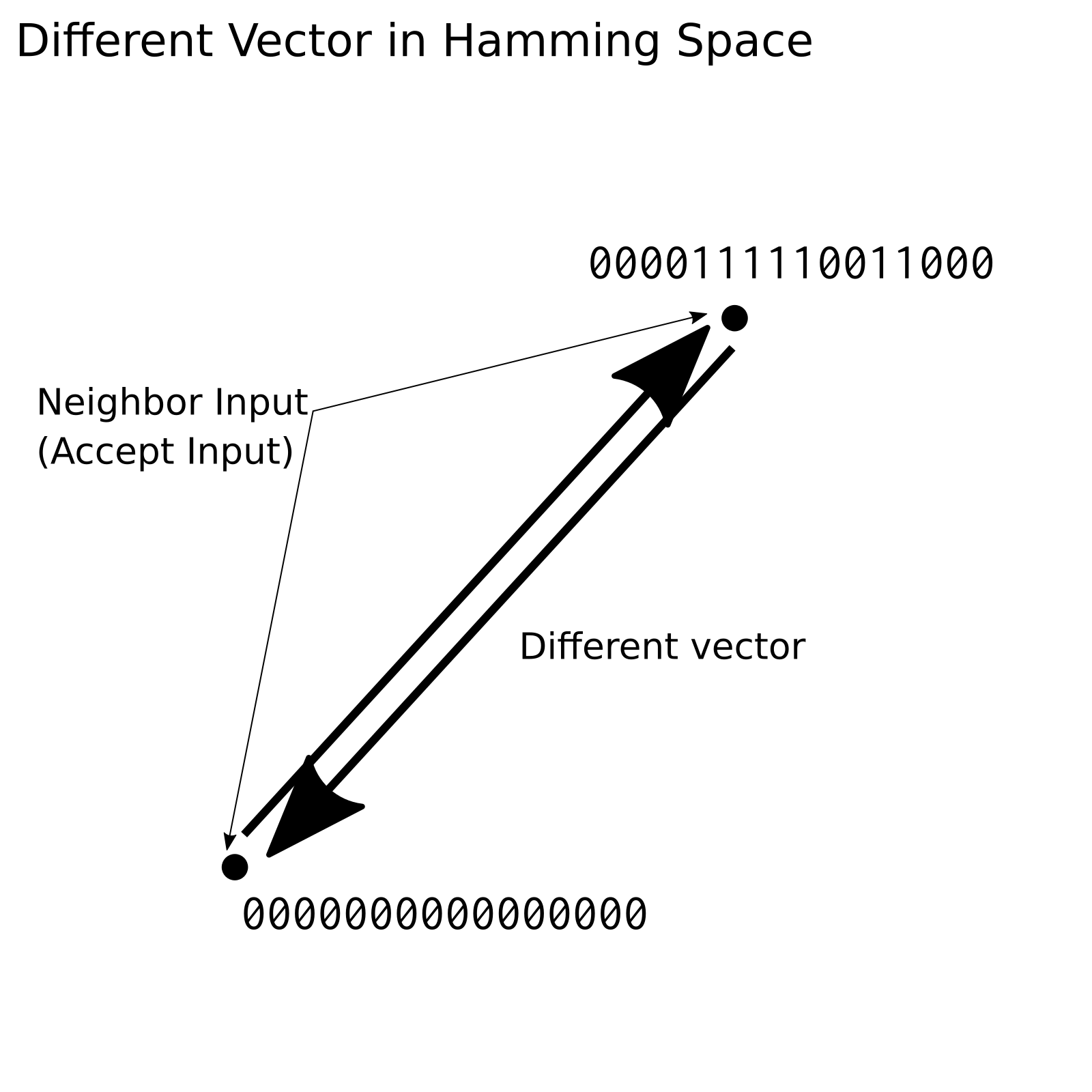}
\par\end{centering}
\caption{\label{fig: Different vector} Different vector}
\end{figure}
``Effective circuit of accept input $t$'' as one of minimal sub
circuit in NNF circuit that decide circuit output as 1 with accept
input $t$. Effective circuit do not include gates which output 0,
or even if these gates change output 0 and effective circuit keep
output 1. 

Figure \ref{fig: Effective circuit} shows example of effective circuit
which circuit is \ref{fig: NNF Circuit} and input is $\left\{ x_{1},x_{2},x_{3}\right\} =\left\{ 1,1,0\right\} $.
Dotted gates do not affect OUTPUT-gate even if the gate negate output,
so effective circuit do not include them.

\begin{figure}
\begin{centering}
\includegraphics{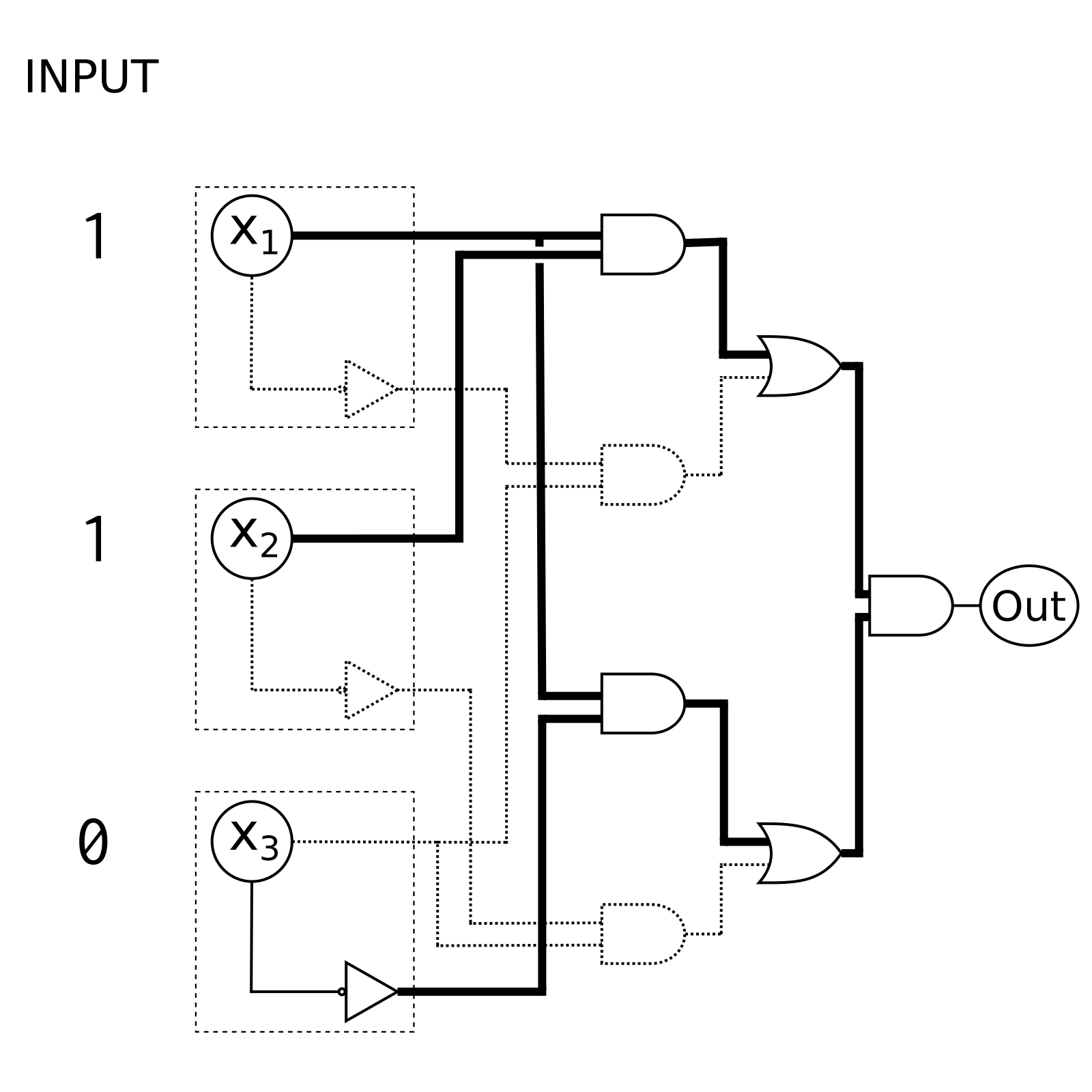}
\par\end{centering}
\caption{\label{fig: Effective circuit} Effective circuit}
\end{figure}
\end{defn}

\begin{thm}
\label{thm: Different variables independency}\textcompwordmark{}

All input variable pair of different variables join at OR-gate in
effective circuit.
\end{thm}

\begin{figure}
\begin{centering}
\includegraphics{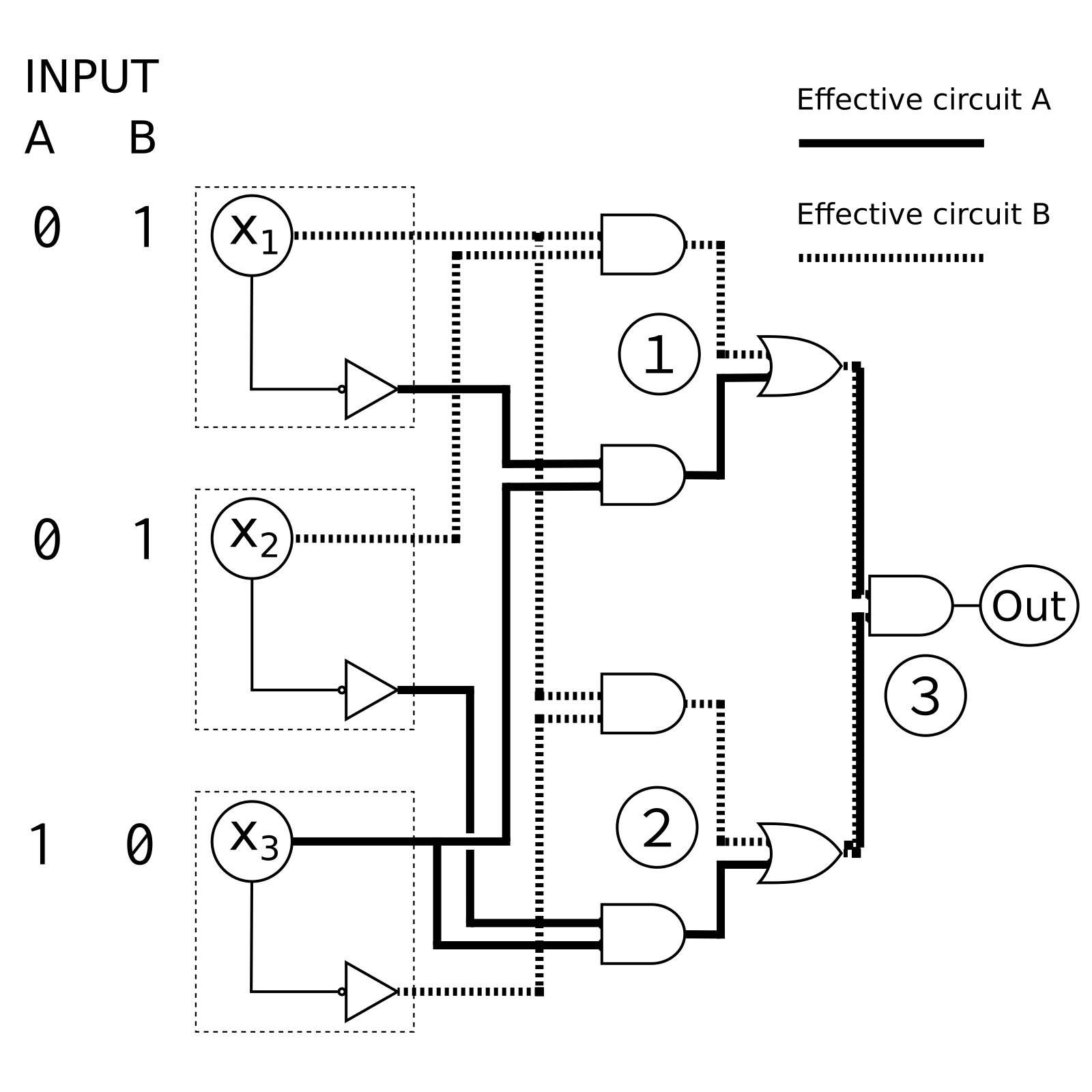}
\par\end{centering}
\caption{\label{fig: Different variables pair} Different variable pair}
\end{figure}

\begin{proof}
Because all input variable pair is $\left\{ 01,10\right\} $ and do
not include $11$ in every input. NNF circuit is almost monotone circuit,
so effective circuit have to to join another accept input $\left\{ 01,10\right\} $
at OR-gate to connect OUTPUT-gate.
\end{proof}
Figure \ref{fig: Different variables pair} shows example of effective
circuit which circuit is \ref{fig: NNF Circuit} and input are $\left\{ x_{1},x_{2},x_{3}\right\} =\left\{ 1,1,0\right\} ,\left\{ 0,0,1\right\} $.
Effective circuit include one of input variable pair, and other side
of variable pair do not become 1 in same input. So AND-gate cannot
meet another effective circuit.

\begin{thm}
\label{thm: Gate independency}\textcompwordmark{}

NNF circuit have at least one unique AND-gate which correspond to
different vector to differentiate neighbor input and boundary input.
\end{thm}

\begin{proof}
Mentioned above \ref{thm: Different variables independency}, all
accept input variable pair of different variables join at OR-gate.
Because NNF circuit is almost all monotone circuit, there are a) b)
case to join effective circuits;

a) some partial different variables meet at AND-gate, and join at
OR-gate these AND-gate output, and meet at AND-gate all OR-gate output.
(see \ref{fig: Different variables pair})

b) all different variables meet at AND-gate, and join at OR-gate after
meeting AND-gate. (see \ref{fig: Gate Independency case b)})

\begin{figure}
\begin{centering}
\includegraphics{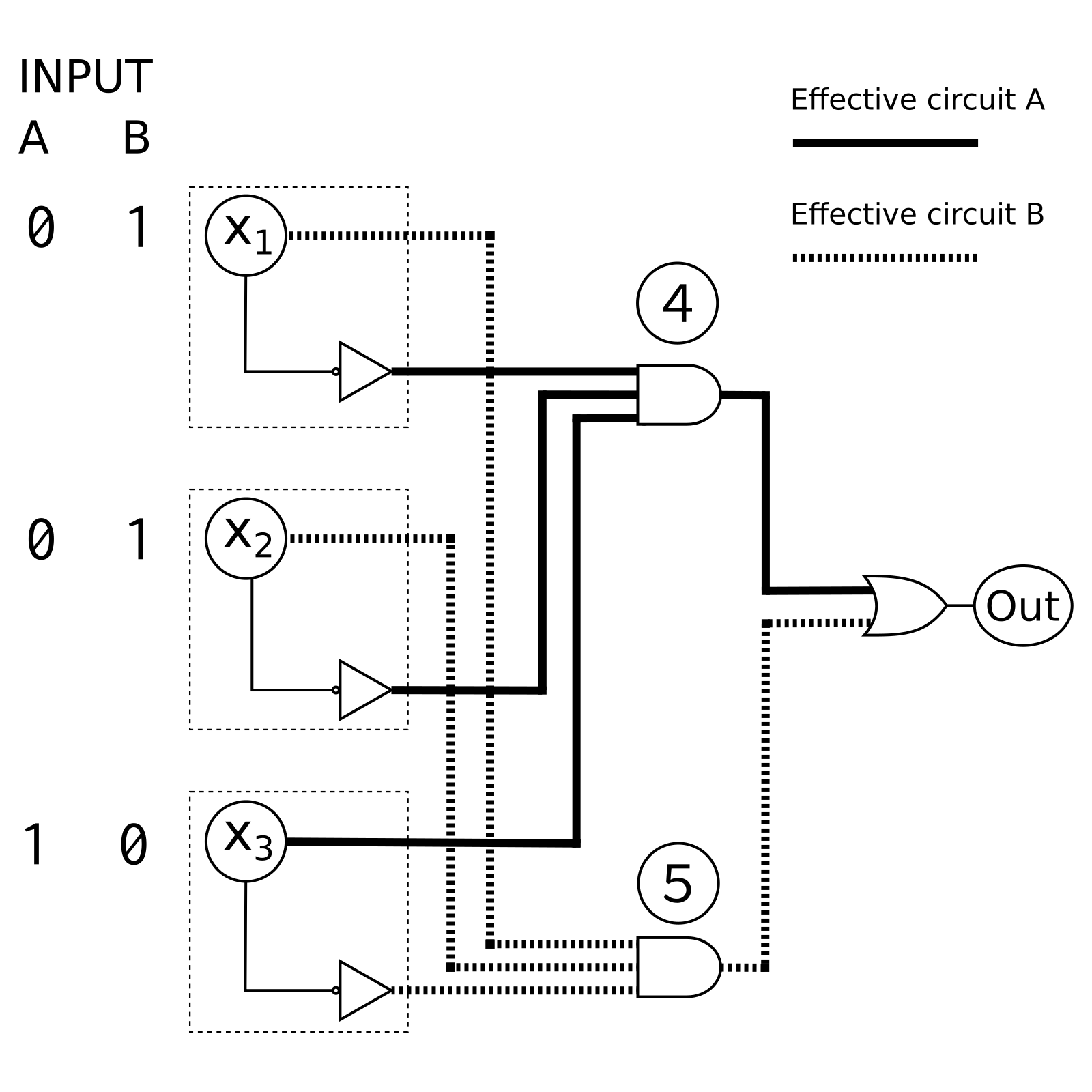}
\par\end{centering}
\caption{\label{fig: Gate Independency case b)} Example of b)}
\end{figure}

Case a), because no boundary input become accept input, some OR-gate
which join different variables become 0 if input is boundary input
(\ref{fig: Different variables pair} 1,2). That is, effective circuit
become 0 if some of these OR-gate become 0, and become 1 if all of
these OR-gate become 1. Therefore, it is necessary that effective
circuit include AND-gate (\ref{fig: Different variables pair} 3)
that meet all these OR-gate which join all different variables. Such
AND-gate become 1 if and only if input include different variables
of one side of neighbor input pair. Each pair of different variables
correspond to different vector, so the AND-gate correspond to different
vector.

Case b), some AND-gate become 1 if and only if input include one side
of different variables. Therefore, trunk of these AND-gate (\ref{fig: Gate Independency case b)}
4,5) does not become 1 if input AND-gate does not include these different
variables. Each pair of different variables correspond to different
vector, so the AND-gate correspond to different vector.

Therefore, NNF circuit have at least one unique AND-gate that correspond
to different vector to differentiate neighbor input and boundary input.
\end{proof}
NNF circuit can emulate DTM in polynomial size, and NNF circuit include
unique AND-gate that correspond to different vector. Therefore, we
can measure problem complexity by counting different vector in problem's
sandwich structure.

\section{Negation HornSAT}

Consider different vector in actual problems. Let consider Negation
HornSAT problem $\overline{HornSAT}$. $\overline{HornSAT}$ can delete
some negative literal which correspond definite clauses. This means
that each $\overline{HornSAT}$ accept input are close each other
in Hamming space. In fact, we can close neighbor distance within constant
distance by devising $\overline{HornSAT}$ description.
\begin{defn}
\label{def: Negation HornSAT}\textcompwordmark{}

We will use the term $\overline{HornSAT}$ as problem if and only
if Horn CNF is $\bot$.

In $\overline{HornSAT}$, we use special description as following; 

$x_{i}$ : Variables in $\overline{HornSAT}$. $i$ in $x_{i}$ is
variable code, and $x$ in $x_{i}$ is constant code. Negative literal
$\overline{x}$ is constant code which length is same as $x$.

$\bot_{i}$ : Disabled Variables in $\overline{HornSAT}$ that $\bot_{i}=\bot$.
$\bot$ in $\bot_{i}$ is constant code which length is same as $x$.

$-$ : Ignored filler code in $\overline{HornSAT}$. 

All another symbol $\wedge\vee()$ are also constant length code which
is same as $x$.
\end{defn}

\begin{thm}
\label{thm: Sandwich structure of Negation HornSAT}\textcompwordmark{}

In $\overline{HornSAT}$, there is some sandwich structure which neighbor
distance is at most constant size, and number of different vector
is at most polynomial size.
\end{thm}

\begin{proof}
Let $t=x_{i}\wedge\left(\overline{x_{i}}\vee\cdots\right)\wedge\cdots\in\overline{HornSAT}$.
can reduce another $t'=x_{i}\wedge\left(\bot_{i}\vee\cdots\right)\wedge\cdots\in\overline{HornSAT}$
because we can delete all literal $\overline{x_{i}}$ by using definite
clauses $x_{i}$. Neighbor distance between $t,t'$ is constant because
difference between $\overline{x_{i}}$ and $\bot_{i}$ is constant
part of $x,\bot$. Because all $\bot_{i}=\bot$, we can reduce all
$\bot_{i}\rightarrow\cdots\rightarrow\bot_{-}\rightarrow\bot$ by
overwriting $-$ at most constant size in each steps, and each neighbor
distance are at most constant. That is, we can reduce $t'$ to $t''=x_{i}\wedge\left(\bot\vee\cdots\right)\wedge\cdots\in\overline{HornSAT}$
with overwriting constant distance. 

The other hand, we can apply above steps all reduction of negative
literals. When some clauses have no variables like $\left(\bot\vee\cdots\vee\bot\right)$,
we can overwrite any code in formula because the formula is $\bot$.
Therefore, all of $\overline{HornSAT}$ have neighbor input that distance
is at most constant.

Consider number of $\overline{HornSAT}$ different vectors. Let different
distance is constant $k$. Because different distance is $k$, number
of different vector is combination of different variables $\left(\begin{array}{c}
n\\
k
\end{array}\right)$ and combination of variables pair in constant code $2^{k}$.

$\left(\begin{array}{c}
n\\
k
\end{array}\right)\times2^{k}=\frac{n!}{k!\times(n-k)!}\times2^{k}\leq O\left(n^{k}\right)$ 

Therefore we obtain theorem.
\end{proof}

\section{Design high complexity problem}

Consider designing high complexity problem. Mentioned in this paper,
computational complexity correspond to problem structure in Hamming
space, especially number of different vector. Therefore, we can design
high complexity problem by designing high cardinal number of different
vector.

We use Almost perfect nonlinear function (APN function)\cite{Dobbertin}
to design high cardinal number of different vector in $F_{2^{n}}$
vector space. APN function $f$ is;

Different function: $f_{a}:F_{2^{n}}\ni x\mapsto f\left(x+a\right)-f\left(x\right)\in F_{2^{n}}$

and $x$ that $f_{a}\left(x\right)=b$($b$: constant) are at most
2.

In this case, $f_{a}\left(x\right)=f\left(x+a\right)-f\left(x\right)$
is different vector of $f\left(x+a\right)$ and $f\left(x\right)$
, so APN function $f$ have many different vector which number is
half of $x$.
\begin{defn}
\label{def: APNR}\textcompwordmark{}

We will use the term;

``$APNR$'' as problem with input $w$, input length $\left|w\right|$,
APN function $f$; 

$w=uv$, $\left|u\right|=\left|v\right|$, $f\left(u\right)=v$
\end{defn}

\begin{thm}
\label{thm: Complexity of APNR}\textcompwordmark{}

$APNR\in PH$
\end{thm}

\begin{proof}
We can compute $APNR$ with constant alternating Turing Machine (ATM)
by computing following way.
\begin{enumerate}
\item Select minimal polynomial $g\left(x\right)$ as existence.
\item Check polynomial solution $\alpha$ as following;\\
Reject $w$ if $\alpha^{2^{d}-1}\neq1$.\\
Select $\alpha^{r}$ ($0<r<2^{d}-1$) as universal, and reject $g\left(x\right)$
if $\alpha^{r}=1$
\item Construct ANR function $f\left(x\right)$ by using $g\left(x\right)$
and APN power functions, and if $f\left(u\right)=v$ then accept $w$. 
\end{enumerate}
We can compute above procedure in polynomial step with constant alternation.
So $APNR\in PH$.
\end{proof}
\begin{thm}
\label{thm: Cardinal number of different vector}\textcompwordmark{}

$APNR$ have $2^{\left|u\right|-1}$ number of different vector.
\end{thm}

\begin{proof}
$APNR$ different vector $w_{p}+w_{q}$($w_{p}=u_{p}v_{p}$,$w_{q}=u_{q}v_{q}$
) is;

$w_{p}+w_{q}=\left(u_{p}+u_{q}\right)\left(v_{p}+v_{q}\right)$

$=\left(u_{p}+u_{q}\right)\left(f\left(u_{p}\right)+f\left(u_{q}\right)\right)$

Because APN function definition, $u_{p}+u_{q}=a$ then there are at
most 2 $\left(u_{p},u_{q}\right)$ that $f\left(u_{p}\right)+f\left(u_{q}\right)=b$.
That is, even if different vector $u_{p},u_{q}$ become same, there
are many different vector $f\left(u_{p}\right)+f\left(u_{q}\right)$
which depend on $u_{p}=u_{q}+a$ value.

Therefore, $APNR$ have many different vector that depend on $u$
value, and number of $APNR$ different vector is $2^{\left|u\right|-1}$.
\end{proof}

\end{document}